\documentclass[pageno]{jpaper}

\usepackage{fancyhdr}
\usepackage[normalem]{ulem}
\usepackage{algorithm}
\usepackage[noend]{algpseudocode}
\usepackage{graphicx}
\usepackage{textcomp}
\usepackage{xcolor}
\usepackage{authblk}
\usepackage{tikz}

\begin{document}

\author[1]{Yuanwei Fang}
\author[1]{Zihao Liu}
\author[1]{Yanheng Lu}
\author[2]{Jiawei Liu}
\author[3]{Jiajie Li}
\author[4]{Yi Jin}
\author[1]{Jian Chen}
\author[1]{Yenkuang Chen}
\author[1]{Hongzhong Zheng}
\author[1]{Yuan Xie}
\affil[1]{DAMO Academy, Alibaba Group}
\affil[2]{Department of Computer Science, University of Illinois Urbana-Champaign}
\affil[3]{Department of Electrical and Computer Engineering, Cornell University}
\affil[4]{Department of Electrical Engineering, Fudan University}
\affil[ ]{\textit {\{yuanwei.fang, zihao.liu, yanheng.lyh, j.chen, y.k.chen, hongzhong.zheng, y.xie\}@alibaba-inc.com}}

\title{NPS: A Framework for Accurate Program Sampling Using Graph Neural Network}

\date{}
\maketitle

\thispagestyle{empty}

\begin{abstract}
	
With the end of Moore's Law, there is a growing demand for rapid architectural innovations in modern processors, such as RISC-V custom extensions, to continue performance scaling. Program sampling is a crucial step in microprocessor design, as it selects representative simulation points for workload simulation. While SimPoint has been the de-facto approach for decades, its limited expressiveness with Basic Block Vector (BBV) requires time-consuming human tuning, often taking months, which impedes fast innovation and agile hardware development.
This paper introduces Neural Program Sampling (NPS), a novel framework that learns execution embeddings using dynamic snapshots of a Graph Neural Network. NPS deploys AssemblyNet for embedding generation, leveraging an application's code structures and runtime states. AssemblyNet serves as NPS's graph model and neural architecture, capturing a program's behavior in aspects such as data computation, code path, and data flow. AssemblyNet is trained with a data prefetch task that predicts consecutive memory addresses.

In the experiments, NPS outperforms SimPoint by up to 63\%, reducing the average sampling error by 38\%. Additionally, NPS demonstrates strong robustness with increased accuracy, reducing the expensive accuracy tuning overhead. Furthermore, NPS shows higher accuracy and generality than the state-of-the-art GNN approach in code behavior learning, enabling the generation of high-quality execution embeddings.

\end{abstract}

\section{Introduction}
In the Post-Moore's Law era, computer architecture innovation has become a critical approach to ongoing performance scaling in modern microprocessors \cite{borkar2011future,hennessy2019new}. 
The increasing popularity and adoption of the RISC-V instruction set \cite{waterman2014risc} has resulted in many microprocessor designs incorporating novel architectural features \cite{chen2020xuantie,demler2020andes,ditzel2022accelerating}, such as customized instructions and acceleration units, embracing rapid innovations. 
Moreover, this golden age of computer architecture \cite{hennessy2019new} desires fast prototyping and agile hardware development.

\begin{figure}[htb]
	\centering
	\includegraphics[trim=0 0 0 0, scale=0.6]{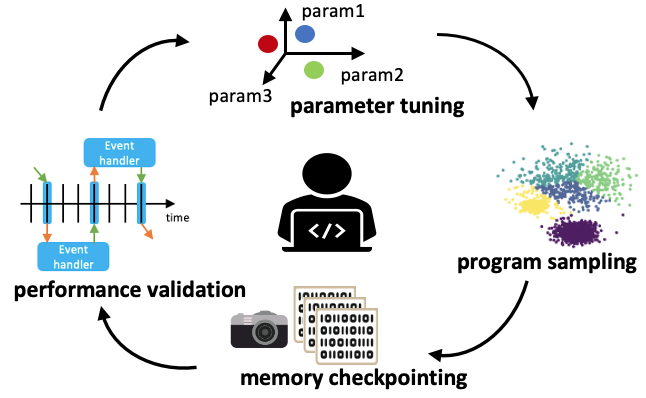}
	\caption{Iterative Accuracy Tuning.}
	\label{figure:iterative_sample}
\end{figure}

Detailed cycle-level simulation is a key tool in computer architecture, but it can't handle large-scale applications, suffering from slow simulation speed. For example, running an application with 1 trillion instructions on a production-grade simulator takes about three years \cite{gem5_tutorial}. To save simulation costs, designers perform program sampling to select representative execution intervals, so-called simulation points, to approximate the entire application for simulation \cite{perelman2003using,hamerly2006using,eeckhout2005exploiting,wunderlich2006statistical,singh2019efficacy}. The accuracy of program sampling has a direct impact on the microprocessor design: it determines the evaluations of various proposed features. However, accurate simulation points are costly, requiring extensive accuracy tuning effort on the settings \cite{perelman2003using,singh2019efficacy}. The standard flow of preparing accurate simulation points is shown in Figure \ref{figure:iterative_sample}, with multiple rounds of sampling, checkpointing, and validation to tweak for accuracy.  
This iterative tuning involves human-in-the-loop and is time-consuming, often taking months.
Historically, large companies rely on the previously built knowledge base and deep experience to reduce the cost of accuracy tuning for their next-generation designs. However, newcomers (e.g., RISC-V startups) without a portfolio on program sampling pay a significant upfront cost. Consequently, the high overhead in accurate program sampling prohibits fast innovation and agile methods in microprocessor development. 

SimPoint \cite{perelman2003using,hamerly2006using} has been the state-of-art program sampling approach for decades. After careful tuning, SimPoint achieves high accuracy on coarse-grained simulation points whose sizes are tens or hundreds of millions of instructions \cite{singh2019efficacy}. However, SimPoint is unstable and lacks accuracy robustness due to its feature vectors' limited resolution in characterizing execution behavior \cite{wunderlich2006statistical}. 
Moreover, designers often resort to small or medium-scale simulation points (e.g., 10K-10M instructions) due to the demand for increased modeling accuracy on low-level simulation \cite{gem5_rtl,shao2016co,verilator,wunderlich2006statistical,armcyclemodel}. For example, hybrid simulation employs a mixed event-based/RTL-level simulation \cite{gem5_rtl,shao2016co}, with fine-grained simulation points for RTL models to account for the slow simulation. Table \ref{table:model_platform} lists the speed and accuracy of common performance evaluation and validation platforms on large CPU designs (billions of gates).
It is challenging for SimPoint to pick small-scale simulation points due to the increased resolution requirement on the feature vectors. In this paper, we focus on accurate program sampling, particularly under fine granularity, with increased accuracy and decreased tuning effort.

\begin{table}[t]
	\centering
	\small
	\begin{tabular}{|p{2.5cm}|p{1.5cm}|p{1.5cm}|p{1cm}|}
		\hline
		Platform & Category & Speed & Accuracy \\
		\hline
		gem5  simulator \cite{gem5paper} & event-based & $\sim$5K inst/sec & medium \\
		\hline
		software transaction simulation \cite{gem5_tutorial,armcyclemodel} & transaction-level & 1K-50K inst/sec & medium \\
		\hline
		FPGA RTL emulation \cite{palladium,karandikar2018firesim} & RTL-level & 50K-1M inst/sec & high \\
		\hline
		software RTL simulation \cite{verilator,ghdl,vcs} & RTL-level & <10 inst/sec & high \\
		\hline
	\end{tabular}
	\caption{Speed and Accuracy of Performance Modeling Platforms.}
	\label{table:model_platform}
\end{table}

The key challenge of accurate program sampling is the design of a high-resolution vector representation for execution characterization, called execution embedding in AI/ML fields. SimPoint uses Basic Block Vector (BBV) \cite{sherwood2001basic} as the execution embeddings.
However, BBV has a few drawbacks in capturing execution behavior. 
First, the code topology information is missing because block connectivity is not encoded. 
Second, BBV is agnostic about instruction semantics, treating instructions with different types and behavior equally. Thus, the similarity between basic blocks is lost. 
Third, BBV can't capture dynamic values. For example, BBV can't distinguish sequential codes with different data-dependent memory accesses. Finally, BBV is defined within a single application without application-wise generality.

\begin{figure}[htb]
	\centering
	\includegraphics[trim=0 0 0 0, scale=0.4]{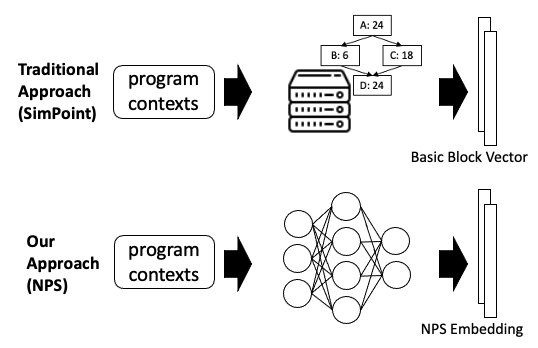}
	\caption{Traditional Approach v.s. Our Approach.}
	\label{figure:landscape}
\end{figure}

We propose Neural Program Sampling (NPS) to address these challenges, producing high-quality execution embeddings for accurate program sampling using a GNN model that learns code behavior. 
Figure \ref{figure:landscape} shows the idea: we train a ``neural microprocessor'' called AssemblyNet, that can execute code using deep learning and let it map execution intervals into embeddings. In contrast, the traditional approach runs on commodity CPUs to profile hand-designed feature vectors (BBV). NPS embeddings capture code topologies such as control-flow and dataflow graphs, instruction semantics, and dynamic contexts of the execution, with cross-application generality. As a result, NPS compensates BBV, enabling code behavior characterization in fine granularity with generalized representations.
Key observations we leverage when applying AI to NPS include:
{\bf 1}. designers are willing to trade machine resources to gain accuracy and reduce the expensive tuning labor, 
{\bf 2}. the embedding generation for benchmark sampling is a one-time effort,
{\bf 3}. checkpointing and accuracy tuning overhead can greatly amortize AI's computational cost, and
{\bf 4}. hierarchical sampling techniques could be developed for a flexible trade-off between runtime cost and sampling accuracy.

In this study, we evaluate NPS on SPEC2006 Integer benchmarks using multiple error metrics. Our results suggest NPS provides robust and accurate program sampling, outperforming the state-of-art SimPoint approach by a large margin. NPS opens up new opportunities in performance modeling, simulation, and benchmarking in microprocessor design. Specific contributions of the paper include:

\begin{itemize}
\item We propose Neural Program Sampling (NPS), a novel framework that provides high-resolution execution embeddings for accurate program sampling, including the description of key designs of graph construction, application tracing, graph snapshot, GNN, and sequence aggregation.
\item We propose AssemblyNet, a GNN graph model and a neural architecture, that learns code execution -- both control-flow and dataflow behavior with high accuracy. 
\item Evaluation of NPS on a diverse set of general-purpose benchmarks.
NPS outperforms the state-of-art approach up to 63\%, achieving an average error reduction of 38\%.
\item Evaluation of accuracy robustness that demonstrates NPS achieves stable accuracy gains under various settings, reducing the need for expensive accuracy tuning.
\item Evaluation of AssemblyNet on a challenging data prefetch task. Compared to the state-of-art GNN approach, AssemblyNet improves the accuracy from 18.2\% to 87.5\%, demonstrating strong capability in capturing code execution.
\end{itemize}

The remainder of the paper is organized as follows. First, Section \ref{background} discusses the background.
In Section \ref{approach}, we describe the NPS framework, including key designs like graph model, neural architecture, and embedding generation.
Then, we evaluate NPS on a variety of general-purpose benchmarks in Section \ref{evaluation} with configurations documented in Section \ref{method}.
Finally, we conclude with a discussion of related works (Section \ref{discussion}) and a summary (Section \ref{summary}).

\section{Background}

\label{background}

\noindent\textbf{Program Sampling} \cite{perelman2003using,wunderlich2006statistical,singh2019efficacy} divides an application into a sequence of fixed-length execution intervals, where a program sampler selects these intervals using feature vectors for a small set of representative simulation points. Thanks to memory snapshotting, simulation points can run on an accuracy validation platform, often a cycle-level software simulator \cite{gem5paper,gem5_tutorial,armcyclemodel}, or an RTL simulation platform \cite{verilator,ghdl,vcs,palladium,karandikar2018firesim}. To tweak for sampling accuracy, designers spend a significant amount of time experimenting with rich sampler settings \cite{perelman2003using,singh2019efficacy} and manual refinements on simulation points. This process is time-consuming and repetitive for multiple rounds. SimPoint \cite{perelman2003using,hamerly2006using} is the state-of-art sampler using K-means clustering and BIC to determine the simulation points.

\noindent\textbf{Basic Block Vector} \cite{sherwood2001basic,perelman2003using} is a popular representation to characterize program execution. In SimPoint, BBV serves as the feature vectors to encode execution intervals, enabling similarity measures. A basic block is a section of
code executed from start to finish with one entry and one exit.
Each dimension represents the entering times of a specific basic block in the application during a certain period.
Code block frequency is a piece of high-level execution information, missing detailed program contexts.
Therefore, BBV captures
how a program changes its behavior over time but in coarse granularity. 

\noindent\textbf{Graph Neural Network}  \cite{hamilton2017inductive,kipf2017semi,velivckovic2017graph} learns how messages are computed from neighbor nodes (Eq. \ref{eq:background_1}), how to aggregate messages (Eq. \ref{eq:background_2}), and how to update node features (Eq. \ref{eq:background_3}) from rich graph data samples. As a result, GNN is not topology-specific and can be generalized to various unseen graphs.
Specifically, a graph can be represented as $G=(V,E)$, where $V$ represents nodes and $E$ represents edges.
In GNN, we denote $E_i$ as the set of edges of type $i$, and $e_{u,v}$ as an edge pointing from node $u$ to $v$. The neighbors of node $v$ of edge type $i$ is defined as $N_i(v) = \{u | (u, v) \in E_i \}$.
We let $h_v^{k}$ denote the feature of node $v$ in layer $k$ ($k = 0$ means the initial node features). We can compute the message that node $v$ received with edge type $i$ from layer $k-1$ in Eq. \ref{eq:background_1}, where $h_u^{k-1}$ is the feature of node $u$ in layer $k-1$ and $f$ is a non-linear function with learnable parameters $\theta_i$. 

\begin{equation}
	m_{i,v}^{k-1}= \sum_{u:N_i(v)}^{} f(h_u^{k-1}, \theta_i)
	\label{eq:background_1}
\end{equation}

Since each edge type has associated learnable parameters, GNN aggregates messages from all edge types (Eq. \ref{eq:background_2}). Here $g$ is the aggregation function (e.g., element-wise summation). 

\begin{equation}
	\widetilde{m}_v^{k-1} = g(m_{i,v}^{k-1} | \textnormal{for all edge type } i)
	\label{eq:background_2}
\end{equation}

Finally, we update the node feature $h_v^{k}$ of layer $k$ with a learnable function $r$.

\begin{equation}
	h_v^{k} = r(\widetilde{m}_v^{k-1}, h_v^{k-1}, \alpha)
	\label{eq:background_3}
\end{equation}

\section{Neural Program Sampling}
\label{approach}
\subsection{Overview}

\begin{figure}[t]
	\centering
	\includegraphics[trim=30 0 0 0, scale=0.47]{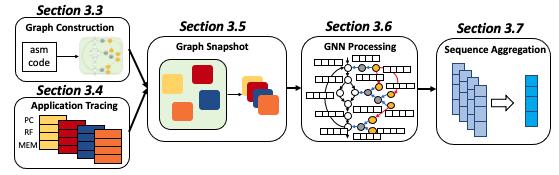}
	\caption{Overview of Generating NPS Execution Embeddings.}
	\label{figure:overview}
\end{figure}

With recent breakthroughs in Graph Neural Network \cite{hamilton2017inductive,kipf2017semi,velivckovic2017graph}, we can model code execution using GNNs, encoding rich contexts like instruction semantics, code structures, and dynamic states along with the instruction stream.
In this paper, we propose a novel program sampling framework, Neural Program Sampling (NPS), based on Graph Neural Network to enable accurate sampling for fine-grained simulation points, shown in Figure \ref{figure:overview}. 
First, NPS constructs an AssemblyNet graph for an application based on its assembly code ({\bf Section \ref{section:graph_model}}).
During runtime, the application's dynamic states and the instruction stream are collected as traces ({\bf Section \ref{section:application_trace}}). Meanwhile, graph snapshots are extracted from the AssemblyNet, incorporating dynamic states ({\bf Section \ref{section:graph_snapshot}}) based on the traces. 
Here, graph snapshots are subgraphs of the AssemblyNet graph with initial node values assigned from dynamic states.
These snapshots are fed into a GNN  module ({\bf Section \ref{section:gnn}}) for graph embeddings, which preloads a trained GNN model ({\bf Section \ref{section:task}} and {\bf Section \ref{section:gnn}}) that can reason about code execution. Finally, NPS aggregates graph embeddings to produce the final execution embeddings for the corresponding intervals ({\bf Section \ref{section:embedding}}). After NPS generates embeddings for all execution intervals, a classic clustering-based sampler \cite{perelman2003using,hamerly2006using} selects the best representative simulation points.

\subsection{Training Task}
\label{section:task}
NPS adopts self-supervised learning when training the GNN model (AssemblyNet). The analogy is the Transfomer \cite{vaswani2017attention}, which is trained with self-supervision by predicting masked words in a sentence, providing word embeddings for downstream NLP tasks. Here, we train AssemblyNet on a data prefetching task that predicts future memory accesses. AssemblyNet doesn't necessarily need annotated training data for this task. Instead, the training data could be easily obtained by the application tracing (Section \ref{section:application_trace}). The data prefetch task allows AssemblyNet to reason and learn about the execution rules, such as code path selection, dataflow propagation, and value computation. For program sampling, NPS drops the task-specific parts and maintains the GNN model to encode execution intervals into graph embeddings.

The data prefetch task requires predicting the correct code path first. The control-flow graph of a program quickly diverges to dozens of code paths after a few memory references. For example, a typical code snippet in {\em 400.perlbench} branches to more than 30 paths only after 5 memory references.
AssemblyNet needs to predict all memory addresses along the code path with all bits correct under the correct order. 
Formally, we define the data prefetch task as follows.
We denote the program representation at instruction $I$ as $S_I$, where $S_I = f(G_I, V_I)$. Here $G_I$ is the graph snapshot (Section \ref{section:graph_snapshot}) and $V_I$ represents the set of dynamic states (Section \ref{section:application_trace}).
We want to train a neural network (Section \ref{section:gnn}) to learn two functions: $f$ and $h$ such that given $G_I$ and $V_I$, we could predict the next $D_I$ consecutive memory addresses following instruction $I$: $h(S_I) = \{addr_1, addr_2, ..., addr_{D_I}\}$, where $addr_i$ is the 64-bit address of $i^{th}$ memory reference.
Success in this task implies the ability to capture fundamentals in code execution, such as resolving branches, activating paths, computing values, and tracking data dependencies. 
For program sampling, we only need $f$ to encode the execution of a program.

\subsection{Offline Construction of AssemblyNet Graph Model}
\label{section:graph_model}

Assembly code is a universal substrate for any high-level programming languages; therefore, AssemblyNet builds the graph model of an application based on it.
The graph model captures code structures and incorporates dynamic states through graph snapshots. It provides a global template for graph snapshots, which are subgraphs fused with dynamic states.
To build the graph model, NPS first creates the backbone according to the assembly code, adding nodes for instructions and connecting them with control flow edges. Moreover, various types of nodes and edges are attached to the graph, representing specific operations and data values for dataflow information.

\begin{figure}[h]
	\centering
	\includegraphics[trim=0 0 0 0, scale=0.26]{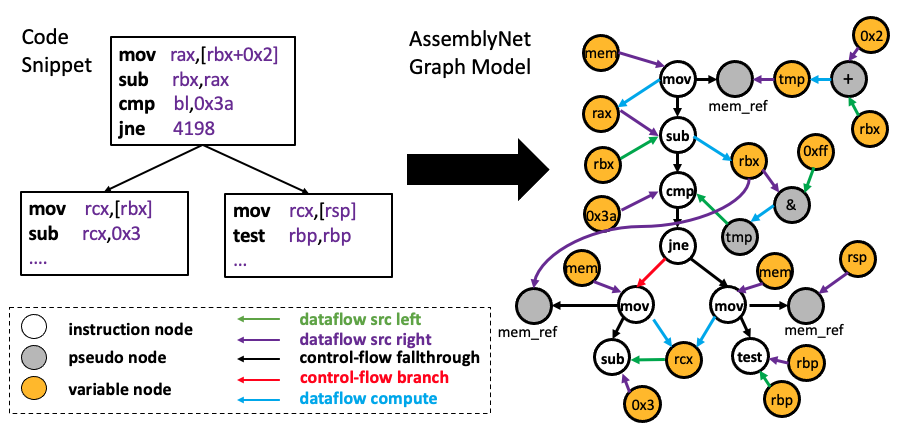}
	\caption{AssemblyNet Graph Model Example.}
	\label{figure:asmgraph}
\end{figure}

AssemblyNet's rationale is to leverage dedicated nodes and edges to capture computation rules, achieving better generality across programs. For example, different edge types are used to distinguish operand position in an instruction, and "bitwise-and" nodes are automatically inserted when dealing with registers with various bit widths (e.g., {\em rax} and {\em eax}). 
Figure \ref{figure:asmgraph} illustrates the graph model of a small code snippet consisting of three basic code blocks and one branch. Next, we discuss the detailed design of the graph model.

\noindent\textbf{Graph Node.}
There are three categories of AssemblyNet nodes: {\em instruction}, {\em pseudo}, and {\em variable}. Each category contains several node types to encode the knowledge of execution rules.
Each {\em instruction} node represents a static instruction in the assembly code, such as {\bf {\em mov}} or {\bf {\em cmp}}. A {\em pseudo} node
represents an operation that happens within an instruction, such as address calculation or memory reference. For example, the instruction {\bf {\em mov rax, [rbx+4]}} which loads the value stored in {\em rbx+4} into {\em rax}, has the source operand associating with an expression that computes {\em rbx+4} followed by a memory load. In such case, two {\em pseudo} nodes are engaged: {\em mem\_ref} and $+$. 
Nodes in the {\em variable} category represent dynamic values, which could be one of the four types: temporary results (e.g., {\em rbx+4}), constants, register values, or values in memory. For example, all yellow nodes in Figure \ref{figure:asmgraph} are {\em variable} nodes.
We summarize the node categories and types in Table \ref{table:node}, documenting their functions, usages, and initialization methods.

\begin{table}[t]
	\centering
	\small
	\begin{tabular}{|p{1.2cm}|p{1.2cm}|p{2.5cm}|p{2cm}|}
		\hline
		category &  type & description & initialization \\
		\hline
		{\em instruction} & {\em inst} & represents assembly instructions (e.g., mov, jne) & token id in the vocabulary\\
		\hline
		{\em pseudo} & $+, -, *$, {\em and}, {\em or}, {\em mem\_ref} & represents operations within an instruction (e.g., memory access) & token id in the vocabulary\\
		\hline
		{\em variable} & {\em tmp}, {\em reg}, {\em const}, {\em mem}& represents a data value (e.g., constant, register, memory) & values from application tracing or zero \\
		\hline
	\end{tabular}
	\caption{Description of Node Categories and Types.}
	\label{table:node}
\end{table}

\noindent\textbf{Graph Edge.}
AssemblyNet has two edge categories: {\em control-flow} and {\em dataflow}, which can be divided into five types to reflect the computation rules (see Figure \ref{figure:asmgraph}).
We use {\em control-flow fallthrough} edges to connect sequential {\em instruction} nodes. If there is a conditional branch diverged from an {\em instruction} node (e.g., {\bf {\em jne}}), we add a {\em control-flow branch} edge to connect the target {\em instruction} node. For example, the {\bf {\em jne}} node has two fan-out edges: branch not taken ({\em control-flow fallthrough}) and branch taken ({\em control-flow branch}). This design facilitates AssemblyNet to distinguish sequential and conditional code, particularly the selection of code paths during execution.
For indirect branches whose destination depends on the runtime register value, AssemblyNet attaches a {\em control-flow branch} edge to each possible destination identified by offline profiling.

There are three edge types under the {\em dataflow} edge category: {\em dataflow src left}, {\em dataflow src right}, and {\em dataflow compute}.
{\em Instruction} and  {\em pseudo} nodes use first two types to connect their data sources. This feature makes AssemblyNet aware of operand locations. Moreover, {\em dataflow compute} edge captures how AssemblyNet computes data of a particular instruction. Nodes with {\em dataflow compute} edges can compute and send outputs to the downstream consumers. In general, dataflow propagation happens along with multiple rounds of message passing in GNN (called GNN layers); values are computed and propagated through {\em dataflow} edges.

\noindent\textbf{Support for Expression.}
An expression calculates the address of memory reference in an assembly instruction. For example,
{\bf {\em movsxd rdx, [rcx+rdx*4]}} loads the value from memory, whose address is calculated by the expression {\em [rcx+rdx*4]}, and stores it to {\em rdx} with sign extension.
For each expression in an instruction, AssemblyNet grows an expression tree recursively, consisting of {\em variable} nodes and {\em pseudo} nodes.
Obviously, the root of an expression tree is always a {\em mem\_ref} node.
There is a {\em control-flow fallthrough} edge that connects the associated {\em instruction} node to the expression root. In such a way, the expression tree could receive the liveness information from the edge: the expression tree is active if the corresponding {\em instruction} node is active.
Moreover, to account for the semantic of value loading from memory, there is always a {\em mem} node (one type of {\em variable} node) connected to the expression tree (Figure \ref{figure:asmgraph}), which is later initialized during runtime.

\noindent\textbf{Graph Construction.}
We describe the AssemblyNet's graph model construction process.
First, NPS disassembles the application binary with {\em gcc} to obtain the
assembly code. Then, NPS forms the graph backbone by adding {\em instruction} nodes and {\em control-flow} edges.
Furthermore, NPS grows the graph with {\em variable} nodes and expands instructions with expressions.
To incorporate the data dependency information, NPS marks the producer instructions of the source registers of each instruction. 
Each source register might have multiple producer instructions because code paths can fork and merge to different targets. The situation becomes complicated when there are loops: an instruction that appears earlier in the assembly code may depend on another instruction that appears later if they are both in a loop. 
To tackle this problem, we develop an algorithm that tracks the propagation of each instruction.
To mark the propagation, we perform a breadth-first search for each instruction $I$ that updates a register $R$, traversing all {\em instruction} nodes dominated by $I$ in the graph. 
The graph traversing process remains active until a new instruction is found to overwrite register $R$. In addition, each instruction $K$ that reads from register $R$ updates its bookkeeping $U_K$ by appending the tuple $(R, I)$: $U_K = U_K \cup \{(R, I)\}$.
We repeat this process until all tuples have been added.
Finally, we traverse the graph again, growing expressions and {\em dataflow} edges based on the bookkeepings.
Figure \ref{figure:asmgraph} shows the graph model generated from the code snippet after running this algorithm.

\subsection{Application Runtime Tracing}
\label{section:application_trace}
The graph model encodes the program's static information, such as code structures, instruction semantics, and data flows.
To capture dynamic contexts, NPS performs application tracing to collect the runtime states such as register and memory values. The application tracing module uses the standard instrumentation tool (e.g., Intel Pin \cite{luk2005pin}) to record the microprocessor architecture states during runtime, such as the program counter and register values, as well as memory references (address and data). The tracing module records the states of every dynamic instruction and saves them in a trace file. Finally, the trace file is sent to the downstream modules for graph snapshotting and GNN processing (Figure \ref{figure:overview}). It also serves the data labels for the training task.

\begin{algorithm}[h]
	\caption{The Graph Snapshot Algorithm}
	\begin{algorithmic}[1]
		\Require $n_{cur}$ \Comment{The current {\em instruction} node}
		\Require $AsmGraph$ \Comment{The AssemblyNet graph model}
		\Require $V$ \Comment{Program's current dynamic states}
		\Procedure{Graph\_Snapshot}{$n_{cur}, AsmGraph, V$}
		\State $G \gets$ empty graph
		\State $T_{access} \gets \{d |$ \# of memory accesses in all longest code paths from $n_{cur}$ without cycles$\}$
		\State $max\_allow\_access \gets$ min($T_{access}$)
		\State $Insts \gets \{v |$ all {\em instruction} nodes in $AsmGraph$ s.t. $D_{v, n_{cur}} \leq max\_allow\_access \}$
		\State $visited \gets {\bf set}(Insts)$, $queue \gets {\bf list}(Insts)$  \Comment{Initialize the bookkeepings}
		\ForAll{$v$ in $Insts$}
		\State $G$.add\_node($v$)  \Comment{Add {\em instruction} nodes first}
		\EndFor
		\While{$queue \not= \emptyset$}  \Comment{Expand the graph snapshot $G$}
		\State $n_i \gets queue$.pop()
		\ForAll{$n_j$ in $Neighbor(n_i)$}
		\If{$n_j \in$ $visited$} {\bf continue}
		\ElsIf{$n_j$ is {\em instruction} node \textbf{and} $n_j \notin Insts$} {\bf continue}
		\ElsIf {$n_j$'s {\em instruction} node $v_j \notin Insts$} \Comment{$n_j$ is {\em variable} or {\em pseudo} node}
		\State {\bf continue}
		\EndIf
		\State $G$.add\_edge($n_i$, $n_j$)
		\State $G$.add\_node($n_j$)
		\State $visited$.add($n_j$), $queue$.push($n_j$)  \Comment{Update the bookkeepings}
		\EndFor
		\EndWhile
		\State  $G \gets$ node\_initialize($G, V$)  \Comment{See Section \ref{section:gnn}}
		\State {\bf return} $G$
		\EndProcedure
	\end{algorithmic}
	\label{algo:extract}
\end{algorithm}

\begin{figure*}[htb]
	\centering
	\includegraphics[trim=0 0 0 0, scale=0.47]{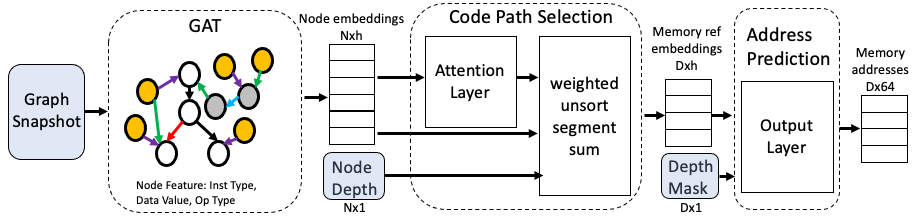}
	\caption{AssemblyNet Neural Architecture (Inputs: Graph Snapshot, Node Depth, and Depth Mask).}
	\label{figure:neural_arch}
\end{figure*}

\subsection{Graph Snapshot Runtime Generation}
\label{section:graph_snapshot}
We explain the graph snapshot algorithm in this section.
Prior work \cite{shi2019learning} suggests the code behavior concerning the current instruction $I$ is determined by the nearby surrounding instructions rather than the entire GNN program graph. AssemblyNet exploits this observation by extracting subgraphs from AssemblyNet and fuse them with dynamic values traced from execution. 
Graph snapshots capture static information such as code topologies, instruction semantics, and dataflow information, and dynamic contexts such as register and memory values during runtime.
To let AssemblyNet reason about multiple consecutive memory accesses (thus more powerful), we define two constraints about the topology of a graph snapshot. {\bf Constraint 1}: Graph snapshot must be a directed acyclic graph in terms of {\em instruction} nodes and {\em control-flow} edges. {\bf Constraint 2}: The number of memory references $D$ must be the same for any code path originating from the current instruction $I$.
Briefly speaking, Constraint 1 simplifies the AssemblyNet learning by excluding loop iteration from consideration. Instead, NPS uses multiple snapshots to represent the repetitive execution behavior of loop iterations. Constraint 2 aligns candidate code paths, each of which might have a different number of memory accesses.
Algorithm \ref{algo:extract} briefly describes the graph snapshot algorithm.
Let $T_{access}$ be the set of the number of memory accesses in all longest code paths $P$ originating from the current {\em instruction} node $n_{cur}$ without any cycles ({\bf Constraint 1}).
$P$ could be obtained by using a DFS (Depth-First Search) traversal with loop detection.
Moreover, we denote $D_{u,v}$ as the minimal number of memory references in all code paths between node $u$ and node $v$.
The $max\_allow\_access$ of the graph snapshot $G$ is the minimum of $T_{access}$. We drop all the remaining instructions in any code path to guarantee $max\_allow\_access$ memory references. As a result, all code paths from $I$ have the same number of memory references ({\bf Constraint 2}). 
We traverse the graph model ({\bf line 5}), collecting all {\em instruction} nodes that satisfy the $max\_allow\_access$ requirement into a list $Insts$.
Finally, we perform a BFS-like traversal, expanding the subgraph to cover as much context as possible by
adding nodes and edges to $G$ ({\bf line 9-18}). After the topology is determined, dynamic values are fused into $G$ ({\bf line 19}) via node initialization (Section \ref{section:gnn}).

\subsection{AssemblyNet Neural Architecture}
\label{section:gnn}
This section discusses the neural architecture of AssemblyNet. We train it with a data prefetch task that predicts consecutive memory reference addresses, given a graph snapshot. 
The {\em code path selection} and the {\em address prediction} are specifically designed for the training task.
We then apply the trained GNN model for generating execution embeddings for program sampling. Figure \ref{figure:neural_arch} shows the overall neural architecture.

\noindent\textbf{GNN Model.}
AssemblyNet uses the Graph Attention Network (GAT) \cite{velivckovic2017graph} to model the behavior of program execution.
GAT assumes the contributions of neighboring nodes are different, deploying a self-attention mechanism to learn the relative weights of messages. Specifically, each node embedding in the current layer is computed from the incoming weighted messages and its node representation from the previous layer. During training, GNN produces $N$ node embeddings ({\em dimension} = $h$) ready for the downstream modules such as code path selection and address prediction.
During runtime, graph snapshots are fed into the GAT model to characterize execution intervals, where each graph snapshot is mapped to a graph embedding via a {\em readout} function.

\noindent\textbf{Code Path Selection.}
Given the current instruction $I$, predicting the code path that $I$ will take is challenging. There are tens of candidate code paths within, for example, five memory references from $I$. In the AssemblyNet's graph model (Section \ref{section:graph_model}), we have dedicated task nodes ({\em mem\_ref}) for address prediction, where each code path has exactly $D$ task nodes. On average, we have $D = 6$ with more than 30 possible code paths (Section \ref{section:eval_asmnet}).
Meanwhile, the neural network must figure out the correct path before it can select the right set of task nodes.
AssemblyNet solves this challenge by applying an attention layer, aggregating the $i^{th}$ ($i \in D$) task nodes from all potential code paths, with weights to compute the most likely $i^{th}$ task node. 
Specifically, the attention layer learns a weight for each task node using its node embedding. Then, embeddings of all potential task nodes are multiplied with these weights.
The weighted embedding of each task node is routed to the specific location according to the {\em Node Depth} (Figure \ref{figure:neural_arch}). {\em Node Depth} specifies the routing destination for the task nodes of $i^{th}$ memory access, indicating the order of memory accesses. Nodes with the same depth are sum-aggregated (e.g., {\em unsorted\_segment\_sum} in TensorFlow \cite{abadi2016tensorflow}). 
Finally, we use the node masking technique \cite{mishra2020node} ({\em Depth Mask}) to accommodate varied numbers of task nodes across graph snapshots, masking out invalid task nodes. We limit the max of $D$ to 20.
The attention mechanism in AssemblyNet essentially selects the most likely code path following instruction $I$ by picking the task node at each memory access (depth).

\noindent\textbf{Address Prediction.}
The code path selection module selects the most likely path using attention, producing the embeddings to predict the address sequence.
We feed the computed ``memory reference" embeddings ({\em dimension} = $D*h$) into the address prediction module, where an MLP transforms these embeddings into 64-bit memory addresses ({\em dimension} = $D*64$). We let the loss function as the masked summation
of sigmoid cross-entropy loss on all $D$ memory addresses.

\noindent\textbf{Node Initialization.}
To incorporate dynamic states, AssemblyNet applies the Neural Code Fusion (NCF) technique \cite{shi2019learning} that assigns dynamic values to graph nodes during runtime, where we modify the register value assignment to better capture code execution invariants.
In NCF, register value assignment is uniform for all {\em variable} nodes under the same register identifier (e.g., {\em rax}). 
However, some of the values should only be computed during GNN message propagation.
As a result, AssemblyNet only assigns register {\em variable} nodes with no incoming {\em dataflow} edges, where other {\em variable} nodes are initialized to zeros, waiting for GNN's computation.
For {\em instruction} and {\em pseudo} nodes, they are initialized with token values in the vocabulary, which are then converted to initial
node embeddings via a learned embedding layer.

\subsection{Embedding Generation}
\label{section:embedding}
Program sampling divides the target application into consecutive execution intervals (e.g., 1M dynamic instructions). We generate an NPS embedding for each such interval.
During embedding generation, AssemblyNet creates a sequence of graph snapshots, where each snapshot covers a short period (usually 50 dynamic instructions). These graph snapshots are mapped to node embeddings and later transformed to graph embeddings by a {\em readout} function. The {\em readout} function could be a permutation-invariant function, such as {\em summation} or {\em mean}, or a more sophisticated
graph-level pooling \cite{zhang2018end,ying2018hierarchical}. In NPS, we simply use the {\em summation}.

To support long instruction sequences, NPS conducts sequence aggregation, summarizing graph embeddings into an execution interval. We experimented with different sequence aggregation algorithms such as mean-aggregation and sequence autoencoder \cite{choi2020encoding}. Autoencoder contrasts mean-aggregation by retaining high-level temporal information when compressing sequences. However, autoencoders are computationally more expensive.
In NPS, we combine both mean-aggregation and autoencoder to support large-scale intervals. 
NPS first averages "subsequences" of instructions, and then applies autoencoder to capture the temporal information of the averaged behavior among "subsequences". This technique helps NPS generate the execution embedding for an interval with more than 100M instructions.
Finally, NPS sends all execution embeddings to a classic clustering-based sampler \cite{hamerly2006using} for representative simulation points.

\section{Methodology}
\label{method}

\subsection{Experiment Setup}
\noindent\textbf{Evaluation Platform.} We use gem5 \cite{gem5paper} as the evaluation platform to get detailed performance statistics. 
Meanwhile, we enhanced gem5 to report the statistics for each execution interval during the simulation.
In the experiments, we use the default O3 model and two calibrated commercial CPU models for evaluation: an open-source Haswell gem5 model \cite{akram2019validation} and a Skylake model calibrated with the Intel Xeon 8163 CPU. 

\noindent\textbf{Workload.} We use the SPEC2006 Integer benchmark \cite{spec2006,spec2006desc} to evaluate NPS. Choosing SPEC2006 enables a direct apple-to-apple comparison on both SimPoint \cite{perelman2003using,hamerly2006using} and NCF \cite{shi2019learning}, the state-of-art approaches for program sampling and code execution learning, respectively. 
The scales of the entire SPEC2006 Integer applications range from 300B to 3800B instructions.
Due to the overhead of detailed cycle-level simulation and NPS processing, we run each application with the reference inputs on 20 billion instructions (240B instructions in total for 12 applications). Although a region with 20B instructions is only a fraction of the application, it is sufficiently large for evaluating high-resolution program sampling with fine-grained simulation points.

\noindent\textbf{GNN.} We use a 4-layer Graph Attention Network (GAT) \cite{velivckovic2017graph} with hyperparameters as follows: node dimension of 256, eight heads, five edge types, $tanh$ as the activation function, vocabulary size of 300, and the hidden layer size of 256. We train AssemblyNet using the Adam optimizer \cite{kingma2014adam} with a fixed learning rate of $10^{-5}$. For training and testing, we partitioned the graph samples collected from 200M instructions with 70\% for training and 30\% for testing. The entire training takes less than 20 hours on Nvidia Tesla V100. After the model is trained, we deploy it to our prototype system to generate NPS embeddings for hundreds of billions of instructions.

\noindent\textbf{System Prototype.} Our prototype system runs on 10 Intel Xeon Platinum 8163 servers. Each server has 48 CPU cores running on 2.5GHz with 800GB of memory. We use CentOS 7 for the operating system, Docker 19 for deployment, and TensorFlow 2.2 \cite{abadi2016tensorflow} for training and embedding generation.

\subsection{Metric}
\label{section:metric}
Program samplers (e.g., SimPoint) select cluster centroids as the simulation points to approximate the original application. As a result, CPI errors are inevitable and could be positive or negative. 
We use two CPI error metrics to measure the program sampling accuracy: {\em Mean Absolute Percentage Error} (MAPE) and {\em Mean Error} (ME).
MAPE is a popular error metric in ML, accumulating the absolute error of each data point (Eq.\ref{eq:1})  \cite{de2016mean}. Since MAPE takes the absolute value for the CPI error of each interval, it is conservative in capturing both +/- errors. {\em Mean Error} (Eq.\ref{eq:2}) \cite{hamerly2006using} simply averages all errors, which is optimistic because it cancels positive and negative errors during the error accumulation. {\em Mean Error} is often used in coarse-grained program sampling but it lacks sufficient error resolution if we need fine-grained program sampling. In the evaluation, we present both error metrics.
To evaluate AssemblyNet's capability, we use data prefetch accuracy as the metric. Note that a correct prediction in data prefetch of multiple consecutive accesses must have bitwise-accurate 64-bit addresses which appear in the right order.

\vspace*{-0.06in}
\begin{equation}
MAPE = \frac{1}{n} \sum_{i=1}^{n} \frac{\left| CPI_{c(i)}-CPI_i \right|}{ CPI_i }
\label{eq:1}
\end{equation}
\begin{equation}
ME = \left|\frac{\sum_{i=1}^{n}CPI_{c(i)}-CPI_i}{\sum_{i=1}^{n}CPI_i }\right|
\label{eq:2}
\end{equation}
where $n$ is the total number of intervals, and $c(i)$ is the cluster centroid that $interval_i$ belongs to.

\section{Evaluation}
\label{evaluation}
In this section, we first demonstrate NPS's program sampling performance. Then, we study the representational benefit of NPS embeddings, followed by an accuracy robustness analysis. Finally, we evaluate the effectiveness of AssemblyNet.

\subsection{Program Sampling Accuracy}
\label{section:psa}
\begin{figure}[htb]
	\centering
	\includegraphics[width=0.95\columnwidth]{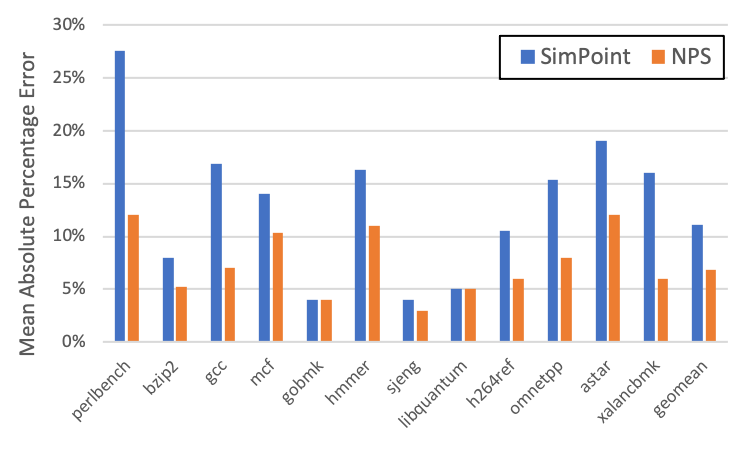}
	\caption{Program Sampling Performance (MAPE).}
	\label{figure:sampling_mape}
\end{figure}

We apply NPS embeddings to program sampling and compare it with the de-facto SimPoint \cite{perelman2003using,hamerly2006using} approach.
NPS learns the execution behavior based on the information from program code structures, instruction types, dataflow graphs, control-flow activeness, and dynamic data values, encoding richer execution context than SimPoint's BBV.
NPS and SimPoint deploy the same sampling algorithm (K-means with BIC \cite{perelman2003using}) but have different vector representations characterizing execution intervals: NPS embedding and BBV, respectively.
We run {\em SPEC2006 Integer} applications with the reference inputs on Skylake gem5 CPU on randomly picked program regions (Section \ref{method}). Each execution interval contains 10 million dynamic instructions. We limit the MaxK to 20 for the sampler to select the simulation points for each application.
Figure \ref{figure:sampling_mape} shows the {\em Mean Absolute Percentage Error}  (Eq.\ref{eq:1}) on 12 applications. MAPE preserves all sampling errors, enjoying higher error resolution than the {\em Mean Error} metric (Eq.\ref{eq:2}). In Figure \ref{figure:sampling_mape}, we can see NPS achieves better accuracy than SimPoint on all 12 applications, achieving an average MAPE of 6.8\% compared to 11\% of SimPoint -- 38\% of sampling error reduction.
In many applications, NPS beats SimPoint by a large margin, more than 50\% and up to 63\% (483.xalancbmk).

\begin{figure}[htb]
	\centering
	\includegraphics[width=0.95\columnwidth]{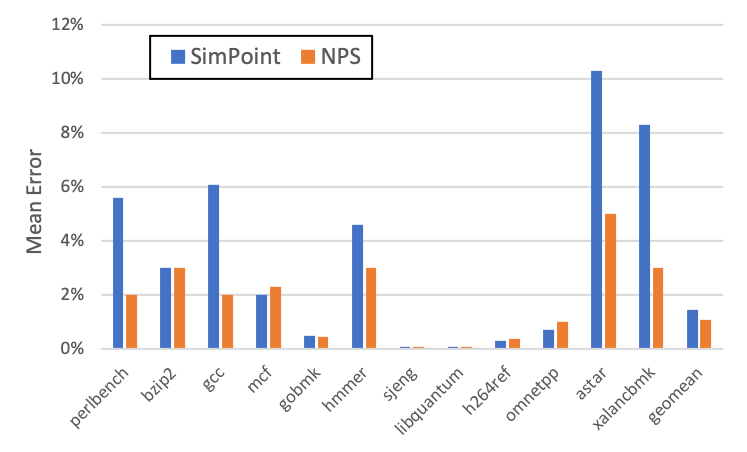}
	\caption{Program Sampling Performance (Mean Error).}
	\label{figure:sampling_me}
\end{figure}

On the other hand, Figure \ref{figure:sampling_me} compares the sampling accuracy between NPS and SimPoint using the {\em Mean Error} metric. Many applications experience tiny sampling errors (less than 2\%) for both NPS and SimPoint. However, since positive and negative errors cancel each other (Section \ref{section:metric}), {\em Mean Error} is coarse-grained and optimistic, underestimating sampling errors. 
As a result, errors can't be fully captured by the {\em Mean Error} metric, which also introduces accuracy instability on the selected simulation points.
Nevertheless, NPS still outperforms SimPoint under {\em Mean Error}, achieving an average error of 1.1\%.
The evaluation results of both metrics suggest that NPS facilitates accurate program sampling by applying high-quality execution embeddings.

\subsection{Execution Embedding Visualization}
\begin{figure*}[htb]
  \subfloat[BBV 403.gcc]{%
    \includegraphics[trim=0 0 0 0, scale=0.23]{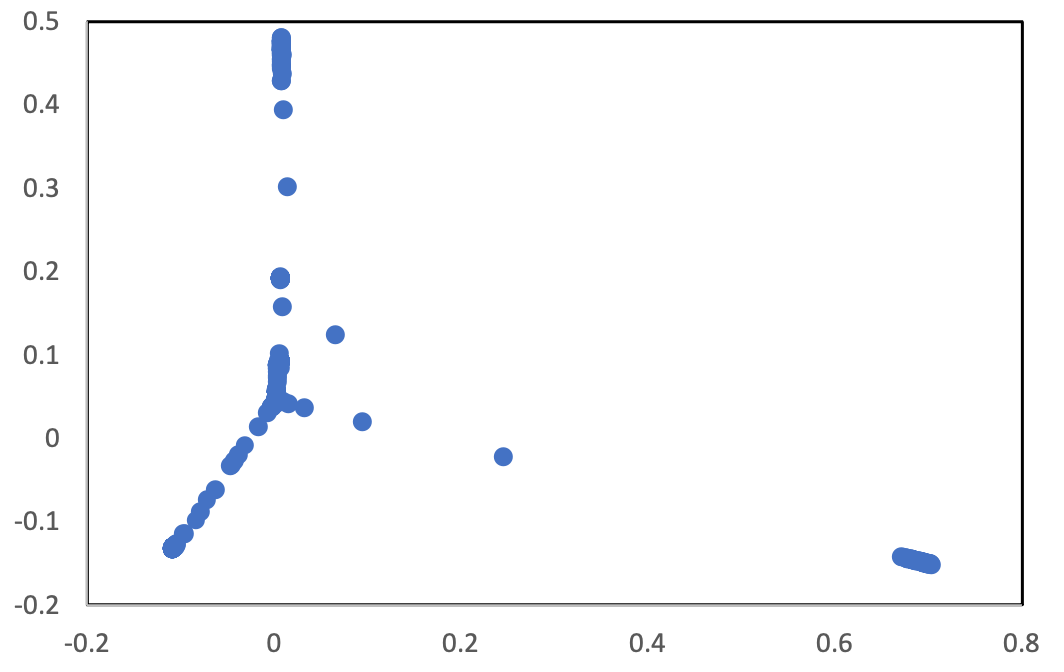}
  }
  \hfill
  \subfloat[NPS 403.gcc]{%
  \includegraphics[trim=0 0 0 0, scale=0.23]{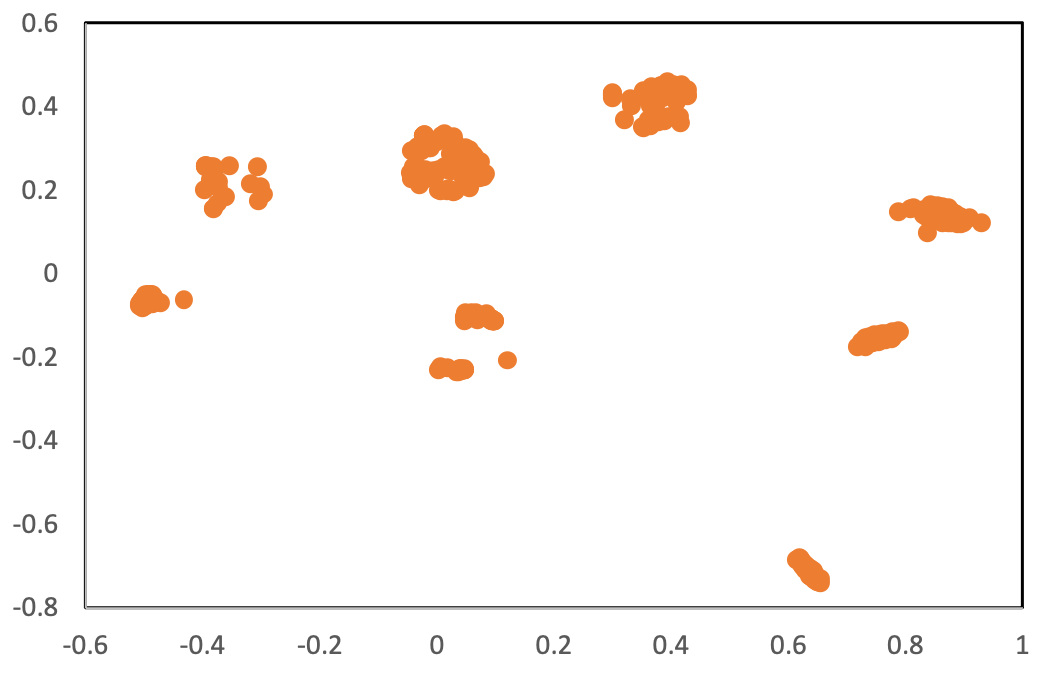}
  }
  \hfill
  \subfloat[BBV 456.hmmer]{%
  \includegraphics[trim=0 0 0 0, scale=0.23]{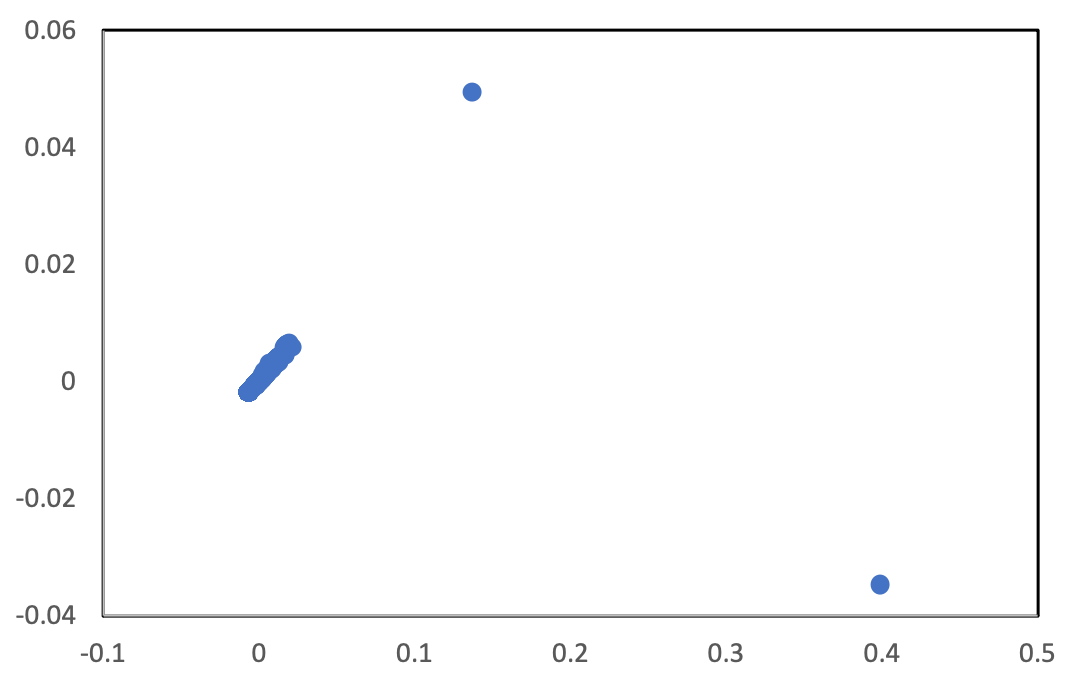}
  }
  \hfill
  \subfloat[NPS 456.hmmer]{%
  \includegraphics[trim=0 0 0 0, scale=0.23]{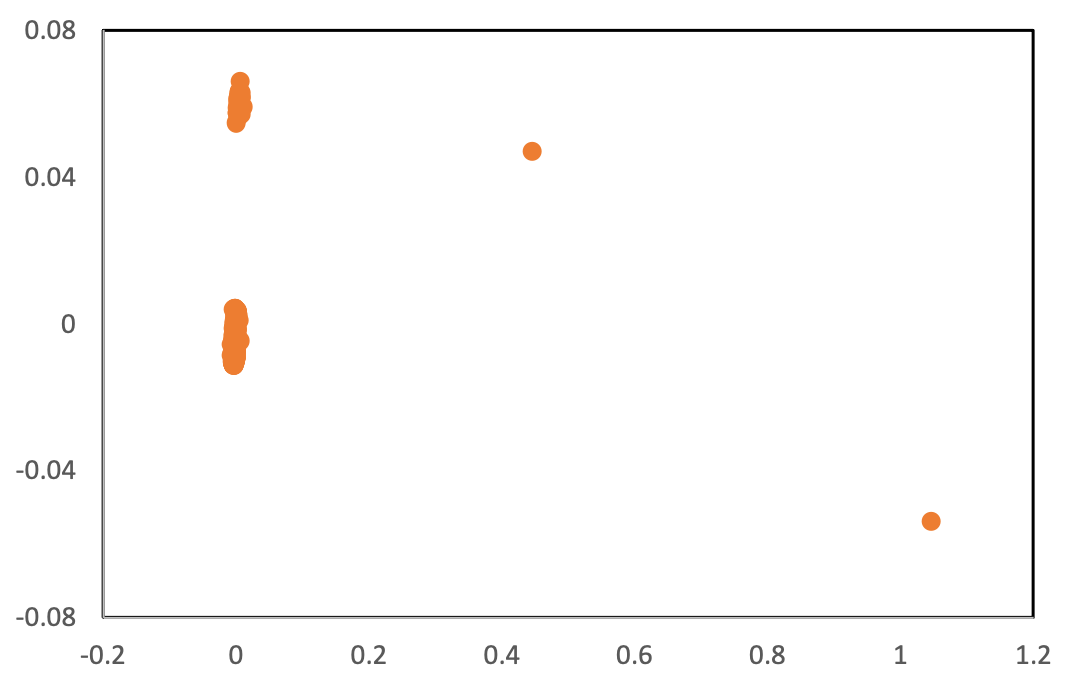}
  \vspace*{-0.05in}
  }
  \hfill
  \subfloat[BBV 473.astar]{%
  \includegraphics[trim=0 0 0 0, scale=0.23]{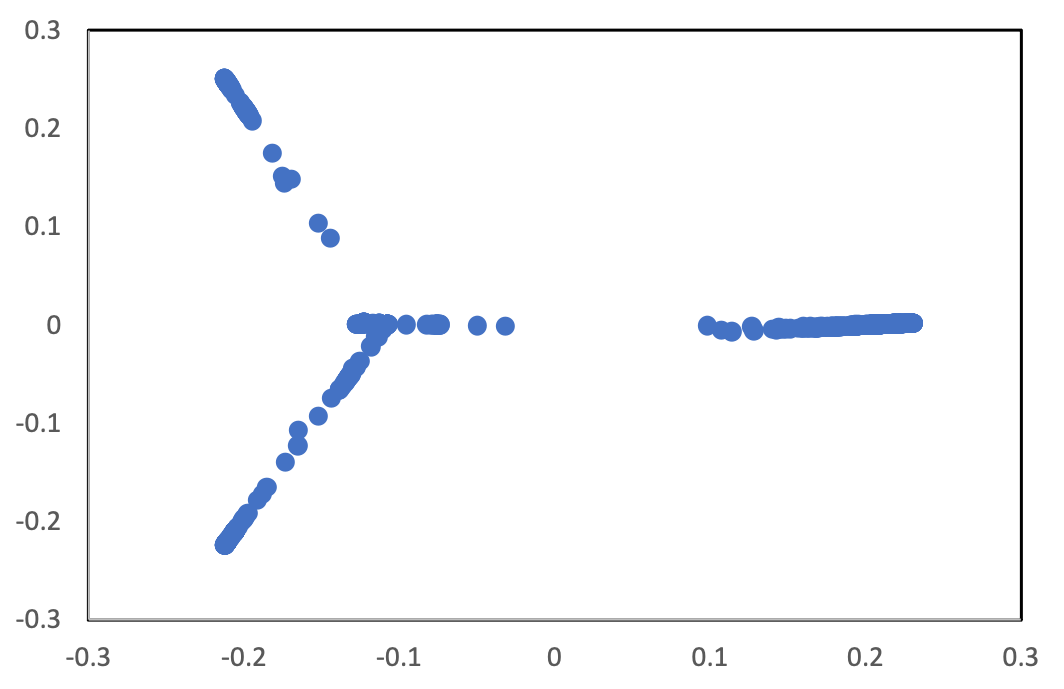}
  }
  \hfill
  \subfloat[NPS 473.astar]{%
  \includegraphics[trim=0 0 0 0, scale=0.23]{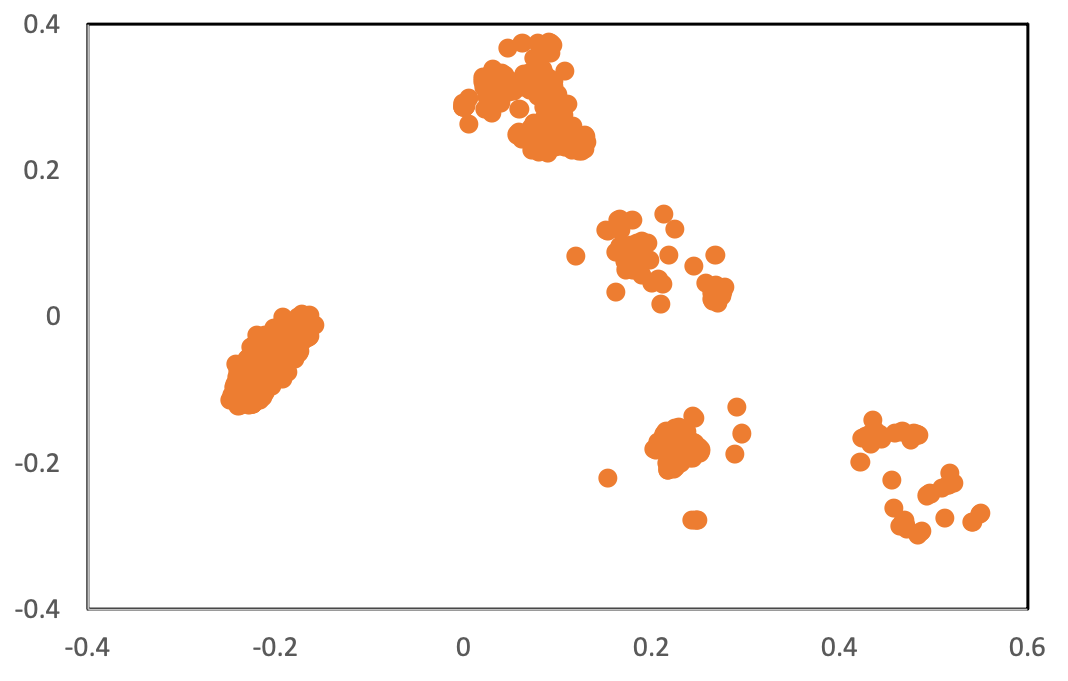}
  }
  \hfill
  \subfloat[BBV 483.xalancbmk]{%
  \includegraphics[trim=0 0 0 0, scale=0.23]{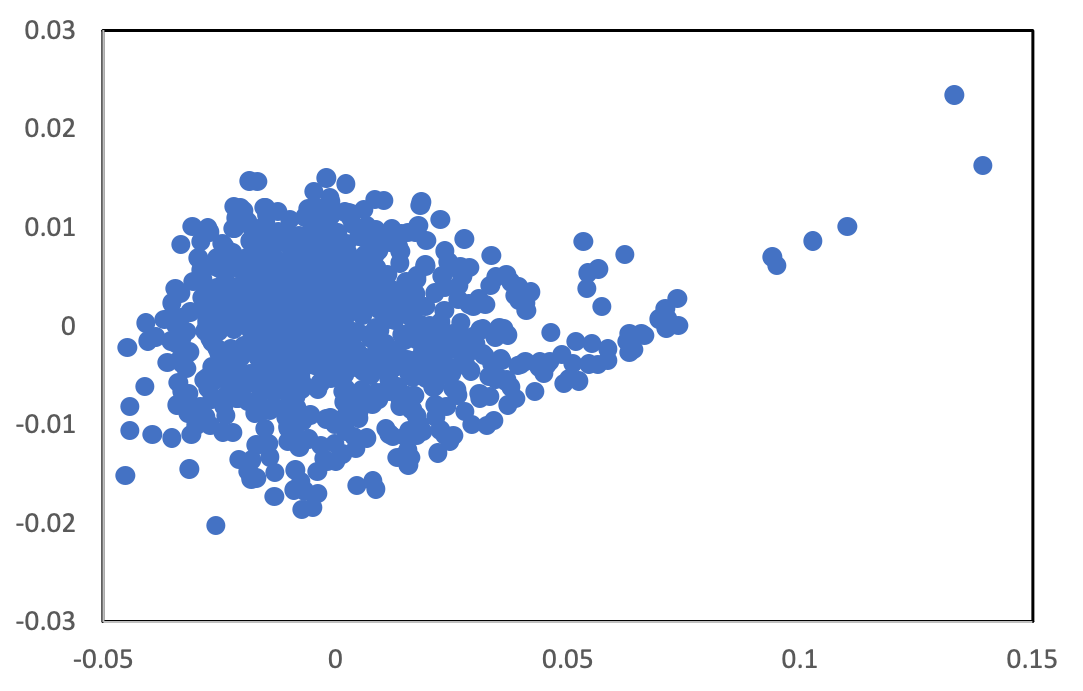}
  }
  \hfill
  \subfloat[NPS 483.xalancbmk]{%
  \includegraphics[trim=0 0 0 0, scale=0.23]{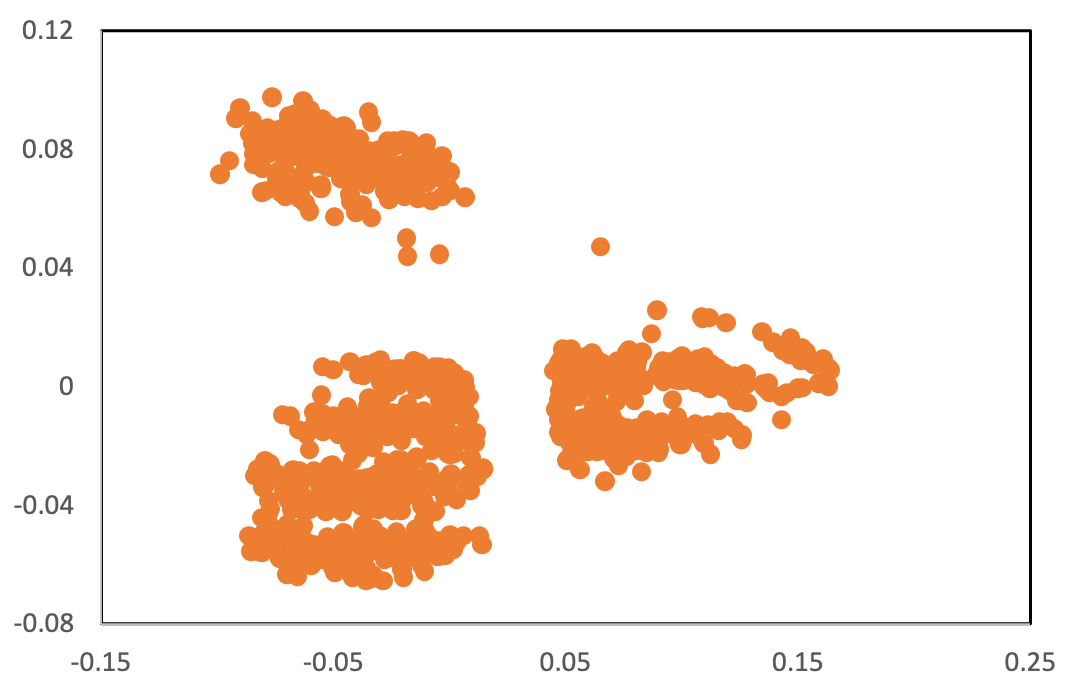}
  }
  \caption{Scatterplot of Execution Embeddings (BBV and NPS).}
  \label{figure:scatterplot}
\end{figure*}

Figure \ref{figure:scatterplot} visualizes the execution embeddings of NPS and BBV on the program regions in Section \ref{section:psa} on four applications.
We leverage the Principal Component Analysis to project execution embeddings into a two-dimensional space. In 403.gcc, we find NPS embeddings enable a better clustering of similar execution intervals. It is because NPS captures features from executions using various information, such as code topologies, instruction semantics, data flow, control flow, and dynamic values, rather than basic block frequencies. BBV distinguishes coarse-grained execution intervals (e.g., 100M instructions) well by comparing different sets of basic blocks and their frequencies. However, under fine-grained intervals (e.g., 500K instructions), the hotspot basic blocks vary slightly, where major differentiation comes from the frequencies. As a result, BBV misses key distinctive features about the execution. Furthermore, BBV leads to fewer degrees of freedom due to the lack of execution information encoded, as shown in Figure \ref{figure:scatterplot}. For example, BBV embeddings spread in 3 directions in 403.gcc. Similar phenomenon pertains to 456.hmmer and 473.astar. In 456.hmmer, BBV and NPS can differentiate two minority sets of execution intervals (see the right part).
However, NPS can further separate the majority of executions into two clusters. While BBV cannot distinguish them due to insufficient context encoded. In 483.xalancbmk, BBV scatters the executions all over the figure -- execution intervals touch various basic blocks, but BBV fails to measure similarities across them. 
Because of BBV's inability to characterize fine-grained execution accurately, SimPoint's accuracy is highly sensitive to the sampler's settings (e.g., {\em seed, MaxK}), which substantially increases the risk of accuracy tuning.
In contrast, NPS utilizes invariants such as code topologies, instruction types, and data values, providing high-quality execution embeddings, which facilitates accurate program sampling.

\subsection{Sampling Accuracy Robustness}
In this section, we study the sampling robustness -- the key to accuracy tuning, varying on simulation resources ($MaxK$ \cite{perelman2003using}), hardware microarchitecture, and sampling granularity.

\begin{figure*}[t]
	\subfloat[SimPoint\label{figure:simpoint_maxk}]{%
		\includegraphics[trim=0 0 0 0 0, scale=0.47]{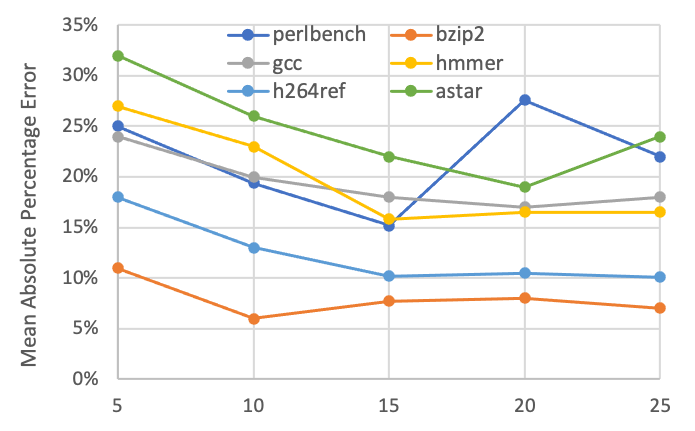}
	}
	\hfill
	\subfloat[NPS (Mean)\label{figure:nps_mean_maxk}]{%
		\includegraphics[trim=0 0 0 0, scale=0.47]{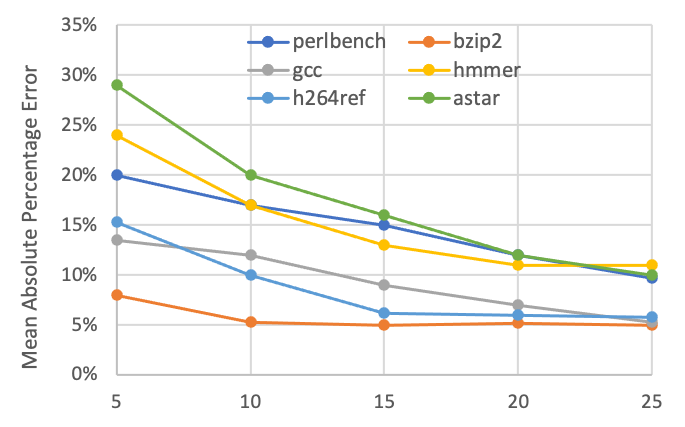}
	}
	\hfill
	\subfloat[NPS (Autoencoder)\label{figure:nps_autoencoder_maxk}]{%
		\includegraphics[trim=0 0 0 0, scale=0.47]{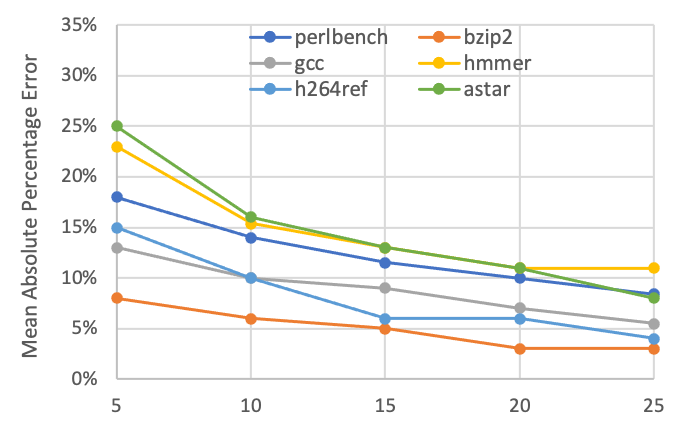}
	}
	\caption{Accuracy Scaling with the Number of Clusters (MaxK).}
	\label{figure:acc_vs_cluster}
\end{figure*}

\noindent\textbf{Scaling accuracy with resources.}
If a sampler allows adding more simulation resources for better accuracy, it scales with simulation resources. 
Good accuracy scalability is essential because engineers often want to trade simulation resources for higher accuracy by adjusting $MaxK$ during the accuracy tuning process.
Additionally, if increasing $MaxK$ doesn't hurt accuracy, the sampling is robust.
In Figure \ref{figure:acc_vs_cluster}, we adjust $MaxK$ and observe the accuracy trend and robustness. Meanwhile, it is up to the sampler to pick any cluster number up to $MaxK$ to maximize the accuracy. Figure \ref{figure:simpoint_maxk} shows SimPoint might have higher errors with larger $MaxK$ such as 401.perlbench. SimPoint's accuracy fluctuates with increased $MaxK$, incurring significant tuning overhead. The root cause is that BBV can't clearly cluster intervals: it misses important execution features and confuses the sampler. 
On the contrary, NPS (Figure \ref{figure:nps_mean_maxk} and Figure \ref{figure:nps_autoencoder_maxk}) is robust, not suffering from reduced accuracy with increased simulation resources.

\begin{figure}[htb]
	\centering
	\includegraphics[trim=0 0 0 0, scale=0.65]{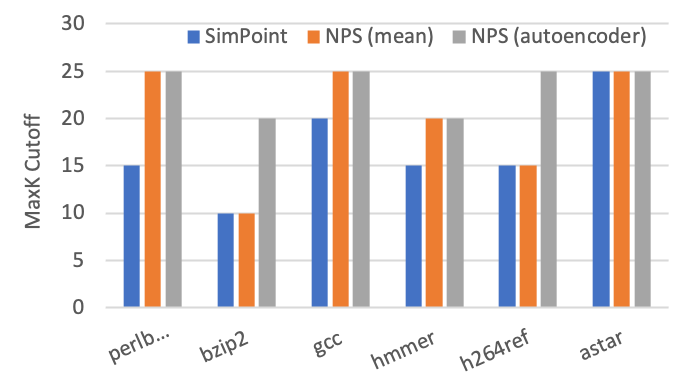}
	\caption{Benefitial MaxK Cutoff.}
	\label{figure:maxk_cutoff}
\end{figure}

Figure \ref{figure:maxk_cutoff} summarizes the beneficial $MaxK$ cutoff, which is the sweet spot in accuracy scaling. Beyond this point, accuracy might hurt or there is no substantial accuracy gain. We find that NPS scales better with a larger $MaxK$ cutoff than BBV.
Moreover, employing an autoencoder in NPS advances the $MaxK$ cutoff in some cases (bzip2 and h264ref).
In summary, thanks to NPS's robustness and accuracy scalability, engineers could be less worried about hurting the accuracy with more simulation points, saving the effort of cumbersome tuning.

\begin{figure}[htb]
	\subfloat[403.gcc]{%
		\includegraphics[trim=30 0 0 0, scale=0.53]{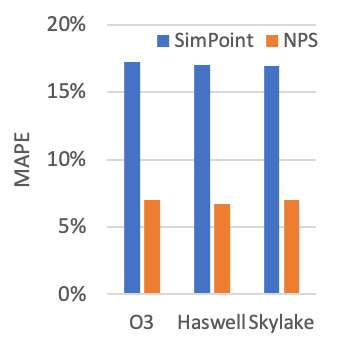}
	}
	\subfloat[456.hmmer]{%
		\includegraphics[trim=0 0 0 0, scale=0.53]{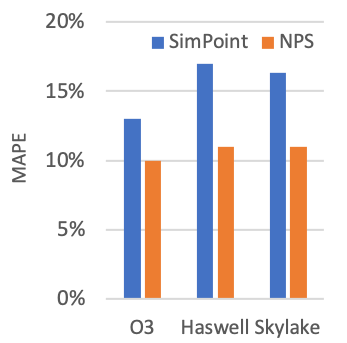}
	}
	\subfloat[471.omnetpp]{%
		\includegraphics[trim=0 0 0 0, scale=0.53]{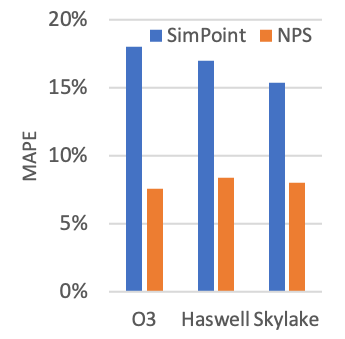}
	}
	\caption{Accuracy Sensitivity on Microarchitecture Variants.}
	\label {figure:arch}
\end{figure}

\noindent\textbf{Microarchitecture Impact.}
Designers perform program sampling to guide their hardware designs, assessing various features, parameters, and configurations based on simulations. Therefore, the selected simulation points must remain accurate on different microarchitectures robustly.
Figure \ref{figure:arch} assesses the sensitivity of sampling accuracy on three applications (403.gcc, 456.hmmer, and 471.omnetpp) on three CPU models described in Section \ref{method}. As we can see, NPS achieves low sampling errors consistently and is robust against CPU models with a variance of less than 1\%.
On the other hand, SimPoint is more sensitive to the underlying hardware platforms, leading to suboptimal simulation points. For example, simulation points vary more than 20\% error rates on the O3 CPU model compared to the Haswell and Skylake models on 456.hmmer.

\begin{figure}[htb]
	\centering
	\includegraphics[width=0.95\columnwidth]{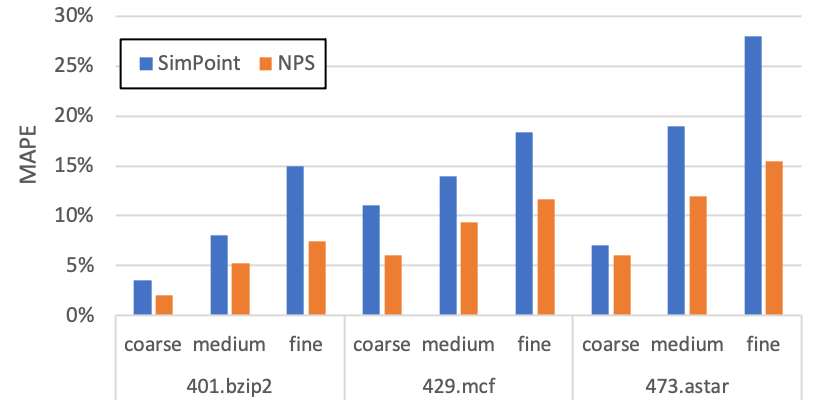}
	\caption{Accuracy Sensitivity on Sampling Granularity.}
	\label{figure:scale}
\end{figure}

\noindent\textbf{Sampling Granularity.}  We study the accuracy robustness in terms of program sampling granularity.
Figure \ref{figure:scale} experiments three levels of granularity: 
{\bf 1}. programs with 2B instructions and intervals of 1M instructions (fine granularity),
{\bf 2}. programs with 20B instructions and intervals of 10M instructions (medium granularity), and {\bf 3}. programs with 200B instructions and intervals of 100M instructions (coarse granularity). This setting mimics program sampling on large-scale applications with coarse-grained simulation points and fine-grained sampling for detailed low-level simulation on small/medium-scale regions. All configurations select the same number of simulation points to keep the sampling complexity the same (equivalent sampling rate). In Figure \ref{figure:scale}, NPS's accuracy gains are robust, outperforming SimPoint in all three levels of granularity on all three applications, thanks to the high-resolution execution embeddings. Moreover, error rates increase significantly in SimPoint when the sampling granularity becomes finer due to BBV's insufficient resolution while NPS experiences better robustness. In short, SimPoint is less satisfied in fine granularity. 
Contrastingly, NPS achieves better accuracy and robustness under various sampling granularities.

\subsection{Evaluating AssemblyNet}
\label{section:eval_asmnet}
\begin{figure}[htb]
	\centering
	\includegraphics[trim=20 0 0 0, scale=0.5]{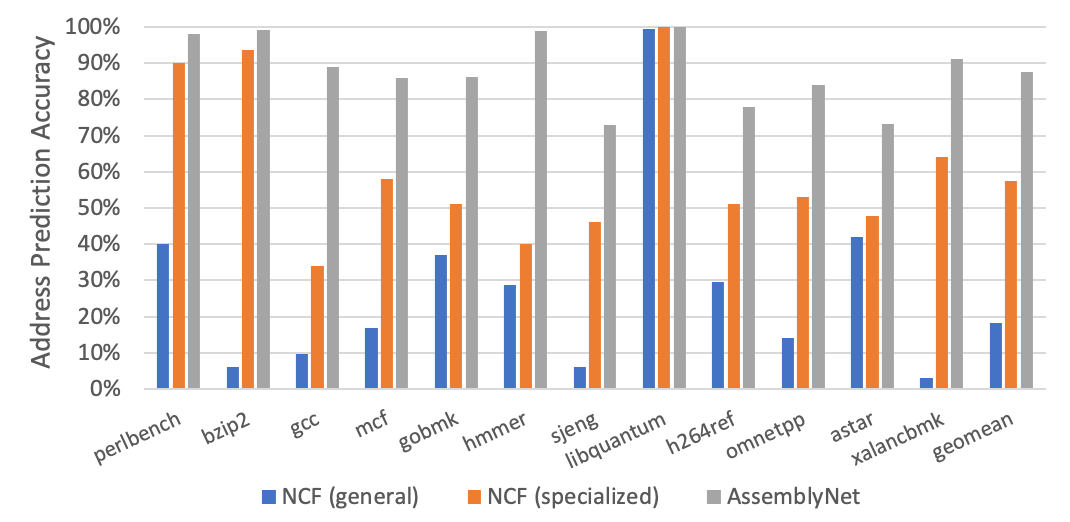}
	\caption{Data Prefetching Task Performance.}
	\label{figure:addr_predict}
\end{figure}

In Figure \ref{figure:addr_predict}, we evaluate the effectiveness of AssemblyNet in learning code behavior via the data prefetch task (Section \ref{section:task}). We use the state-of-art NCF \cite{shi2019learning} approach as the baseline. \textit{NCF (general)} uses a single unified GNN model for all applications. \textit{NCF (specialized)} uses a GNN model (GGNN  \cite{li2015gated}) for each of the applications, which is the design in the original paper \cite{shi2019learning}. For \textit{NCF (general)} and \textit{AssemblyNet}, we use Graph Attention Network (GAT) \cite{velivckovic2017graph} as the GNN model. Comparing these two shows that AssemblyNet achieves better generalization across applications, with an average accuracy of 87.5\% versus 18.2\% of NCF. Regarding \textit{NCF (specialized)}, it can achieve good performance with an average accuracy of 57.4\%. However, it requires training a separate GNN model for each application.
When comparing  \textit{NCF (general)} and \textit{NCF (specialized)}, we find NCF greatly suffers from generalization across applications due to the inability to capture application invariants. As a result, NCF cannot learn code execution across applications in a general fashion.
In contrast, AssemblyNet demonstrates superior performance in accuracy and generality. Moreover, AssemblyNet produces 6x prefetch addresses on average than NCF in a single graph snapshot. Therefore, AssemblyNet is more powerful than NCF in capturing dynamic program behavior, paving the way for generating high-quality execution embeddings.

\subsection{Application-wise Execution Embeddings}

\begin{figure}[htb]
	\centering
	\includegraphics[trim=0 0 0 0, scale=0.28]{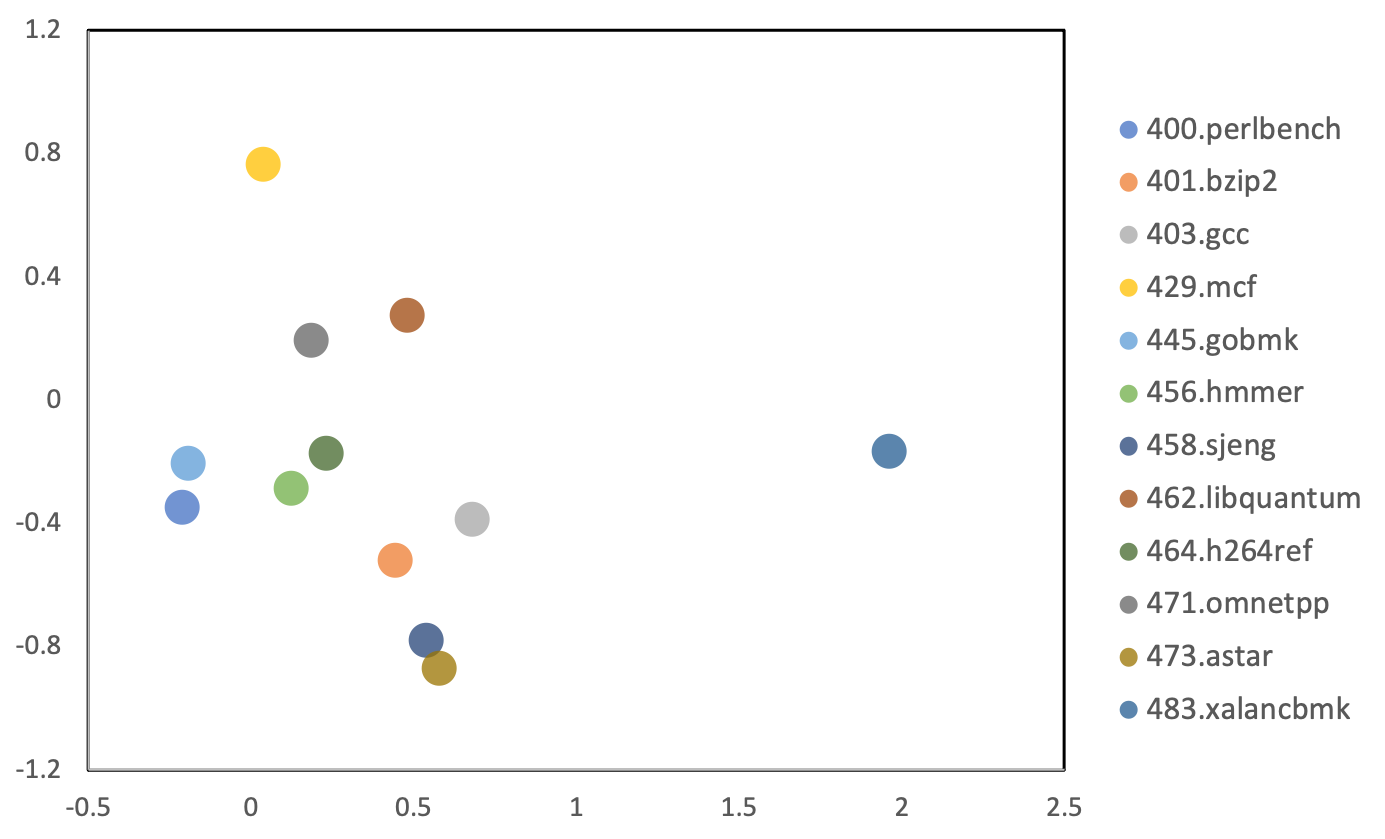}
	\caption{Scatterplot of SPEC2006 Integer Applications.}
	\label{figure:pca_all}
\end{figure}

NPS learns execution invariants, such as dataflow propagation, value computation, control-flow activeness, and instruction semantics. Thus, it can map executions from different applications into the same vector space. Figure \ref{figure:pca_all} shows the scatterplot of the program regions in Section \ref{section:psa} for 12 SPEC applications. Thanks to NPS, we can directly compare executions from different applications, while BBV is limited within a single application due to the non-transferability of knowledge encoded in the basic blocks. In Figure \ref{figure:pca_all}, we can see 483.xalancbmk is quite different from the other applications. In addition, 458.sjeng is very similar to 473.astar, where two applications actually fall into the same \textit{Artificial Intelligence} category \cite{spec2006desc} -- 458.sjeng is about chess AI playing and 473.astar is about game's AI pathfinding. In summary, NPS provides a generalized representation for code execution.

\section{Discussion and Related Work}
\label{discussion}

\subsection{Performance Evaluation for Microprocessor Design}

\noindent\textbf{Program Sampling.}
Large-scale benchmarks are expensive and sometimes infeasible to run on a simulator. Program sampling \cite{perelman2003using,hamerly2006using,eeckhout2005exploiting,wunderlich2006statistical,singh2019efficacy,prieto2021fast,sabu2022looppoint} is a popular statistical approach to reduce the workload simulation cost when evaluating various design choices in a microprocessor. 
The industry standard sampler is SimPoint \cite{perelman2003using,hamerly2006using}, which samples an application with BBV \cite{sherwood2001basic} for representative simulation points.
Next, checkpointing tools \cite{patil2010pinplay,patil2021elfies} snapshot the memory states of each sample so that designers can simulate them in parallel without fast forwarding. The accuracy of the simulation points is validated against the original program, and designers tune the sampler's settings (e.g., {\em seed, maxk, threshold}, etc.)  \cite{singh2019efficacy}, adjusting the selection to tweak the accuracy. This process repeats until the accuracy is satisfied. As a result, accurate program sampling is traditionally time-consuming: performance validation and checkpointing are slow to run, and there are multiple rounds of tuning.
On the other hand, NPS provides a new approach for high-quality execution embeddings, enabling accurate program sampling with reduced tuning overhead.
Due to the fundamental richer contexts encoded, both static and dynamic information, NPS embeddings are more effective than BBV serving as the feature vectors for execution intervals, especially in fine-grained program sampling. 
Moreover, NPS's robustness eliminates expensive accuracy tuning overhead, saving a substantial amount of time for microprocessor designers.

\noindent\textbf{Workload Synthesis.}
Another thread of work in reducing simulation cost is via workload synthesis by constructing microbenchmarks \cite{awad2014stm,badr2020mocktails,chen2020dwt,eeckhout2005exploiting,panda2017proxy}.
Microbenchmarks are carefully hand designed to emphasize particular aspects of the workloads (e.g., memory access patterns) for microprocessor evaluation.
For example, there are locality metrics such as {\em reuse} and {\em stride} distance \cite{awad2014stm,badr2020mocktails,chen2020dwt,panda2017proxy}, and control-flow metrics like instruction mix, branch rate, and instruction dependency \cite{eeckhout2005exploiting,chen2011modeling,panda2017proxy}. However, workload synthesis requires deep human experience of the target workloads and the hand-designed feature engineering, incurring non-recurring engineering cost when creating new microbenchmarks from different benchmark domains. In contrast, NPS provides a generalized approach to sample critical parts of the target workloads from various fields.

\noindent\textbf{Simulation Infrastructure.}
For performance evaluation and validation purposes, simulators are essential tools in microprocessor development, which have a broad spectrum on
speed and accuracy trade-offs.
Software simulators offer great flexibility in modeling OS-capable full-system workloads, either event-driven \cite{gem5paper,gem5_tutorial} or transaction-based \cite{armcyclemodel}.
However, the flexibility and the relatively fast speed come at the cost of accuracy loss.
On the other hand, RTL-level simulation provides accurate performance but suffers from extremely slow simulation if executed in software \cite{verilator,ghdl,vcs}.
As a result, there are hybrid simulators such as {\em gem5+rtl} \cite{gem5_rtl} and gem5-Aladdin \cite{shao2016co} that embed RTL models inside full-system simulation to have a balance between flexibility, simulation speed, and modeling accuracy.
FPGA-based emulations \cite{palladium,karandikar2018firesim} accelerate the RTL modeling but are constrained by the available machine resources. Recently, there have been active research efforts in developing AI-assist simulators \cite{renda2020difftune,li2022simnet} using DNNs. NPS helps designers break down workloads into fine-grained simulation points using AI techniques so that designers can run them cost-effectively on the best suitable simulation platforms.

\subsection{Modeling Software/Hardware System with AI}
\noindent\textbf{Compile Time.}
There is an increasing research interest in modeling compiler's behavior using AI.
For example, compiler researchers apply GNN models to static code analysis such as type inference \cite{wei2020lambdanet}, bug fixing \cite{dinella2020hoppity,allamanis2017learning}, operator scheduling \cite{zhou2019gdp}, and code similarity comparison \cite{alon2019code2vec}. In these tasks, an abstract syntax tree or a computation graph is used as the backbone of the GNN graph model. 
Beyond GNNs, researchers also explore auto-vectorization \cite{haj2020neurovectorizer} with Deep Reinforcement Learning to generate high-performant vectorized code. Like these works, NPS leverages AI and builds a program-specific graph model to capture static information. Additionally, NPS also focuses on encoding runtime contexts for high-quality execution embeddings as well.

\noindent\textbf{Execution Time.}
Various DNN models have been proposed to model control-flow and dataflow code behavior during runtime \cite{allamanis2017learning,bieber2020learning,lym2019branchnet,shi2019learning,hashemi2018learning}. For example, BranchNet \cite{lym2019branchnet} targets learning hard-to-predict branches with a CNN model. 
NCF \cite{shi2019learning} uses GNNs for branch prediction and data prefetching. 
Voyager \cite{shi2021hierarchical} proposes a hierarchical neural data prefetcher. 
NALU \cite{trask2018neural} trains neural networks for arithmetic operations on floating-point numbers. 
Among these works, data prefetch is a popular task \cite{shi2021hierarchical,hashemi2018learning,shi2019learning} that reflects the all-round abilities to learn code execution. 
NPS draws inspiration from NCF. However, we found NCF cannot generalize well across applications, only being effective within a single application.
NPS contrasts NCF with a generalized graph representation (AssemblyNet) across applications, focusing on the learning of invariant computation rules among programs (e.g., operator location, dataflow propagation, control-flow activeness, etc.).
Moreover, NPS's training task is more challenging, combining code path selection and consecutive memory references prediction together, rather than simply predicting the next subsequent memory access \cite{shi2019learning}. Thanks to the difficult task, NPS generates high-quality execution embeddings.

\noindent\textbf{Design Time.}
Deep learning techniques have gained remarkable progress in many aspects during the system design phase. 
For example, researchers propose to model cloud workloads with RNNs \cite{bergsma2021generating} for the next-generation cloud platform design.
Moreover, DiffTune \cite{renda2020difftune} uses AI to learn the parameter settings of a simulator to minimize the simulation errors on a given set of workloads.
SimNet \cite{li2022simnet} builds a neural simulator for cycle-level microarchitecture simulation.
BOOM Explorer \cite{bai2021boom} provides an efficient design space search of RISC-V cores with a Gaussian model and a deep kernel learning function.
Lin et al. \cite{lin2020deep} optimize the loop placement in routerless Network-on-Chip via Deep Reinforcement Learning. Mirhoseini et al. \cite{mirhoseini2021graph} propose an AI tool for chip floor-planning and device place\&route by encoding netlists into GNNs with DRL optimization. Similarly, NPS is an AI-assisted design tool for agile development. NPS provides robust fine-grained program sampling with high accuracy, easing the performance evaluation on microprocessor designs.

\section{Conclusion}
\label{summary}

This paper presents Neural Program Sampling (NPS), a novel framework that produces high-quality execution embeddings using dynamic snapshots of GNN, enabling accurate program sampling for fine-grained simulation points.
In NPS, AssemblyNet learns code execution by training on a data prefetch task. NPS creates graph snapshots and transforms them into execution embeddings for program sampling, which outperforms the de-factor SimPoint approach by a large margin. NPS's capability on generating execution embeddings opens up exciting research opportunities in microprocessor performance modeling, benchmarking, simulation, and evaluation.

\bibliographystyle{plain}
\bibliography{refs}

\end{document}